# Strain-dependent Dynamics of Interactions of Iodine-vacancy pairs in $CsPbI_3$: A DFT and Machine-Learning Molecular Dynamics Study

*Donggeun Kang*[1], *Seongju Lee*[1], *Yonggoon Jeon*[1*]

[1]*Department of Physics and Chemistry, Korea Military Academy, Seoul, 01805, Republic of Korea*

*Corresponding author: 12-10011@kma.ac.kr

Iodine-vacancy migration and association are important factors in the structural stability of halide perovskites. Although neighboring vacancies have been suggested to form energetically favorable pairs, their stability and reconfiguration under finite-temperature lattice deformation remain unclear. Here, we combine machine-learning molecular dynamics (MLMD) and density functional theory (DFT) to investigate neighboring iodine vacancies in tilted $CsPbI_3$ under isotropic and biaxial strain. The MLMD trajectories reveal a nonmonotonic dependence of short-time hopping fractions on strain, with strong isotropic compression favoring point-like separation and tension promoting pair-preserving relay hopping. Notably, biaxial compression produces distinct orientation-dependent responses: vacancy pairs initially aligned with the compressed plane preferentially undergo separation, whereas those perpendicular to the plane predominantly exhibit relay hopping. Complementary static calculations using DFT and the SevenNet-0 potential yield repulsive interactions for all configurations examined. These findings highlight the importance of finite-temperature dynamics and strain-dependent local geometry in understanding vacancy-pair stability and reconfiguration in halide perovskites.

**Key words**: Halide perovskites, Iodine vacancies, Defect interactions, Strain engineering, Machine-learning molecular dynamics, Density functional theory

## 1. Introduction

Inorganic halide perovskites, particularly $CsPbI_3$, have emerged as core materials for high-efficiency photovoltaics and optoelectronic devices [1,2]. However, the migration of intrinsic ionic defects, particularly iodine vacancies ($V_I$), can contribute to operational instability in halide perovskite devices [3]. Lattice strain acts as another critical parameter governing this defect behavior [4,5]. Featuring a mechanically soft ($PbI_6$) octahedral network, perovskites are highly sensitive to external stresses, such as thermal expansion mismatch or mechanical bending [6]. Consequently, structural distortions profoundly modulate defect formation energies and transport properties [7–10].

Previous studies have established a foundation for understanding perovskite defects, elucidating the fundamental physics of isolated point defects and the energetic favorability of specific defect pairs [11]. However, the interactions between neighboring iodine vacancies under charge-neutral conditions remain to be examined in the context of intrinsically tilted lattice configurations. Octahedral tilting is relevant not only to finite-temperature lattice dynamics but also to the intrinsic energetics of $CsPbI_3$, as previous first-principles studies have shown that cubic $CsPbI_3$ is unstable with respect to octahedral tilting and that tilted perovskite configurations have lower static energies than the cubic reference [12,13]. Therefore, the energetic preference between tilted and non-tilted structures should be assessed when selecting the host lattice for defect-interaction analysis. Furthermore, significant progress has been made in understanding the impact of lattice strain on defect behavior. How different strain modes, such as isotropic and biaxial strain, influence the interactions and dynamic evolution of neighboring vacancies remains insufficiently understood. Elucidating how adjacent defects reconfigure under realistic operating conditions is an imperative challenge for controlling device stability.

Here, we combine machine-learning molecular dynamics (MLMD) and density functional theory (DFT) to examine neighboring iodine-vacancy pairs in a representative tilted $CsPbI_3$ structure. We investigate how strain mode and initial pair orientation affect the fractions of trajectories exhibiting pair-preserving relay hopping or point-like separation within a finite observation window. Static DFT interaction energies provide complementary energetic context for these dynamical observations.

**2. DFT and MLFF Computational methods**

DFT calculations were performed using Quantum ESPRESSO 7.5 [14,15] within the Perdew-Burke-Ernzerhof (PBE) generalized-gradient approximation [16], employing projector-augmented-wave (PAW) pseudopotentials [17] from PSLibrary 1.0.0 [18]. Wavefunction and charge-density cutoffs of 80 and 640 Ry, respectively, were employed, higher than those previously used for first-principles calculations of $CsPbI_3$ [19]. Brillouin-zone integration was performed using a converged $4 \times 4 \times 4$ $k$-point mesh for the $2 \times 2 \times 2$ supercell and $2 \times 2 \times 2$ $k$-point mesh for the $4 \times 4 \times 4$ supercell. A Gaussian smearing of 0.005 Ry was applied, and the electronic self-consistency threshold was set to $10^{-7}$ Ry. For defect interaction-energy calculations, lattice vectors were fixed while internal atomic coordinates were relaxed using BFGS algorithm, with a force-component convergence threshold of 0.077 eV/Å. Within each comparison, identical numerical parameters were applied to the pristine, single-vacancy, and vacancy-pair structures. All DFT defect calculations considered charge-neutral conditions. Spin–orbit coupling (SOC) was omitted, following previous studies reporting relatively small SOC effects on defect formation energies in lead iodide perovskites [20].

The SevenNet-0 potential was used for static Machine Learning Force Field (MLFF) calculations in the defect calculations. The applicability of SevenNet-0 to large-scale atomistic simulations has been demonstrated by its ability to reproduce DFT-derived structural characteristics, including radial and angular distributions, in molecular dynamics simulations of disordered systems [21]. Atomic positions were optimized using BFGS while the strained lattice vectors remained fixed. Relaxation was terminated when the maximum atomic force magnitude fell below 0.01 eV/Å, with a maximum of 1,000 optimization steps. Following previous DFT-MLFF validation approaches for iodine migration in $CsPbI_3$ [22], the SevenNet-0 potential was further evaluated against DFT calculations along two representative iodine migration paths (**Fig. S1** in Supplemental Material). The MLMD simulations were performed using the Atomic Simulation Environment (ASE) [23] within the NVT ensemble at 350 K, utilizing a Berendsen thermostat [24] with a relaxation time of 100 fs. The equations of motion were integrated with a 1 fs time step for a total duration of 12 ps per system. This duration consisted of a 2 ps equilibration phase followed by a 10 ps production phase, during which atomic configurations were sampled every 50 fs.

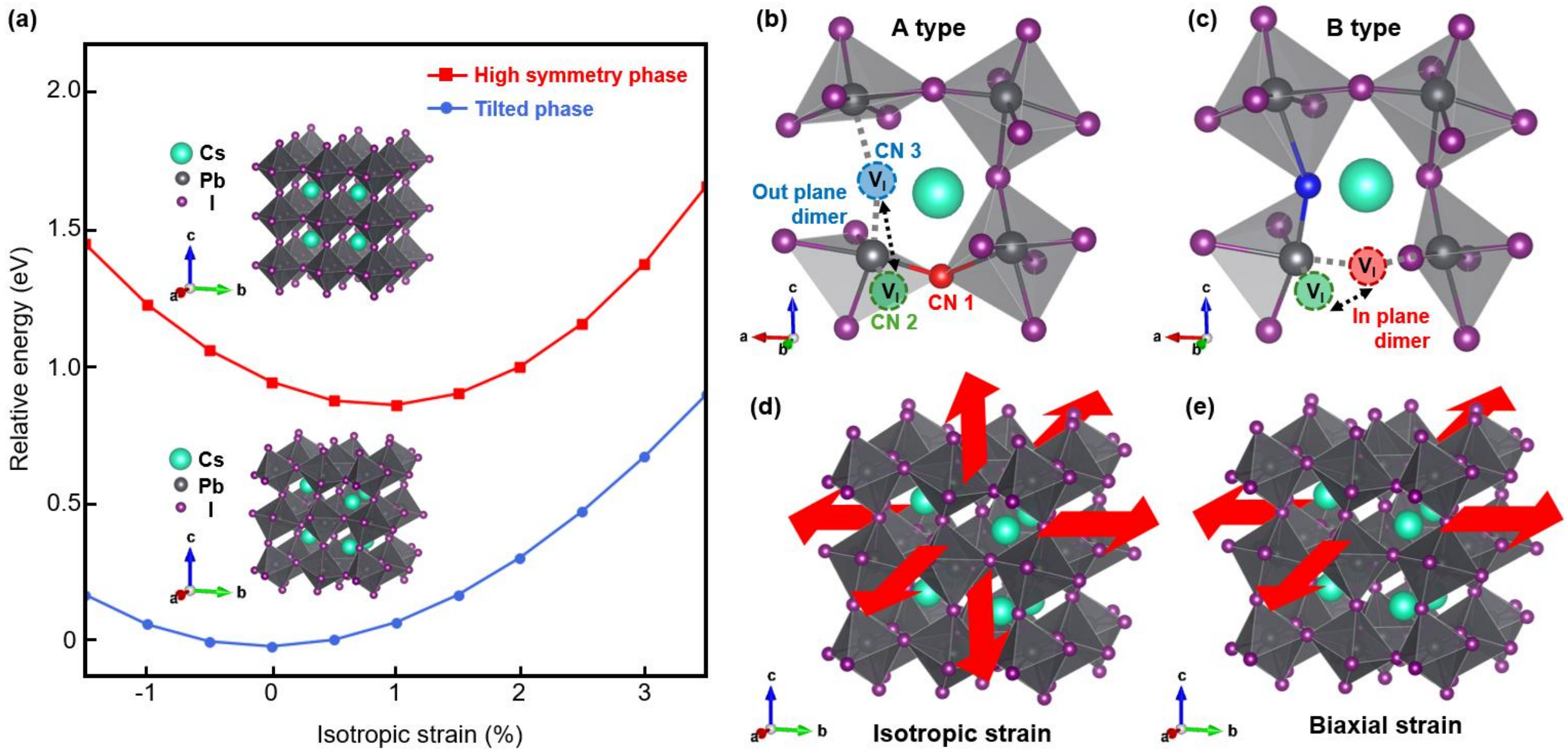

**Figure 1. Structure models and applied strain modes** (a) Relative single-point energies of the high-symmetry and tilted 2×2×2 $CsPbI_3$ structures under isotropic strain (−1.5% to 3.5%), calculated using the SevenNet-0 potential and referenced to the unstrained tilted structure. (b, c) Representative spatial arrangements of iodine-vacancy pairs: (b) the out-of-plane A-type pair and (c) the in-plane B-type pair. Distinct local coordination environments (e.g., CN 1, CN 2, CN 3) adjacent to the vacancy sites are highlighted in the A-type model. (d, e) Schematic representations of the (d) isotropic and (e) biaxial strain modes.

### 3. Structure models and classification of vacancy-pair dynamics

A representative tilted $CsPbI_3$ structure was generated by relaxing symmetry-broken configurations derived from a high-symmetry 2 × 2 × 2 supercell containing 40 atoms. Initial hydrostatic relaxations were performed using the pretrained SevenNet-0 machine-learning interatomic potential without additional fine-tuning [21,25], optimizing both atomic positions and isotropic cell volume while preserving the cell shape and angles. The lowest-energy configuration was selected as the baseline tilted model, and its equilibrium lattice constant was subsequently refined using DFT. As illustrated in **Fig. 1a**, the tilted phase exhibits significantly lower single-point energies, evaluated via SevenNet-0, than the high-symmetry phase across the evaluated isotropic strain range (−1.5 % to 3.5 %), supporting its adoption for subsequent defect calculations. The relaxed 2 × 2 × 2 structure served as the common reference for the subsequent calculations. For the static interaction-energy survey using SevenNet-0, the pristine reference structure was replicated twice along each lattice vector to construct 4 × 4 × 4 supercell containing 320 atoms, increasing the separation between periodic defect images. For MLMD, the pristine reference structure was replicated four times along each lattice vector to construct 8 × 8 × 8 supercell containing 2,560 atoms before vacancy introduction. Vacancies were introduced after replication, rather than replicated together with the host structure.

To examine the dependence of vacancy-pair behavior on local geometry and applied strain, initial iodine-vacancy pairs were constructed by removing two adjacent iodine atoms sharing a common Pb polyhedron. These pairs were categorized into two representative geometries: out-of-plane A-type configurations, oriented predominantly along the **c**-axis, and in-plane B-type configurations, confined within the **ab**-plane (**Figs. 1b** and **c**). Isotropic strain was applied uniformly along all three crystallographic directions, whereas biaxial strain was confined to the **ab**-plane while the out-of-plane cell dimension was held constant (**Figs. 1d** and **e**).

To systematically sample the defect landscape in MLMD, the initial vacancy-pair geometries were enumerated based on the distinct iodine-site environments within the underlying tilted 8 × 8 × 8 supercell (**Fig. 2a**) structural motif. Symmetry-equivalent defect placements across the expanded supercell were clustered into common orbits, yielding 96 unique representative vacancy-pair configurations for each applied strain condition. The same initial-velocity seed was used across the representative configurations without analyzing sensitivity to different initial-velocity realizations. Vacancy dynamics were analyzed by representing each iodine bridge site as the midpoint of its two neighboring Pb atoms and identifying vacant sites based on their distances to the nearest iodine atoms. The identified vacancy sites were tracked across the saved MD configurations, as illustrated by a representative single-vacancy trajectory in **Fig. 2b**. The initial pair orientations were categorized based on the angle, $\theta_z$, defined between the pair axis and the crystallographic **c**-axis. Configurations aligning closely with the out-of-plane or in-plane directions were classified as A-type and B-type pairs, respectively, while intermediate states were designated as ambiguous orientations. The detailed vacancy-tracking algorithm and quantitative orientation-classification criteria are provided in Supplemental Material **S2**.

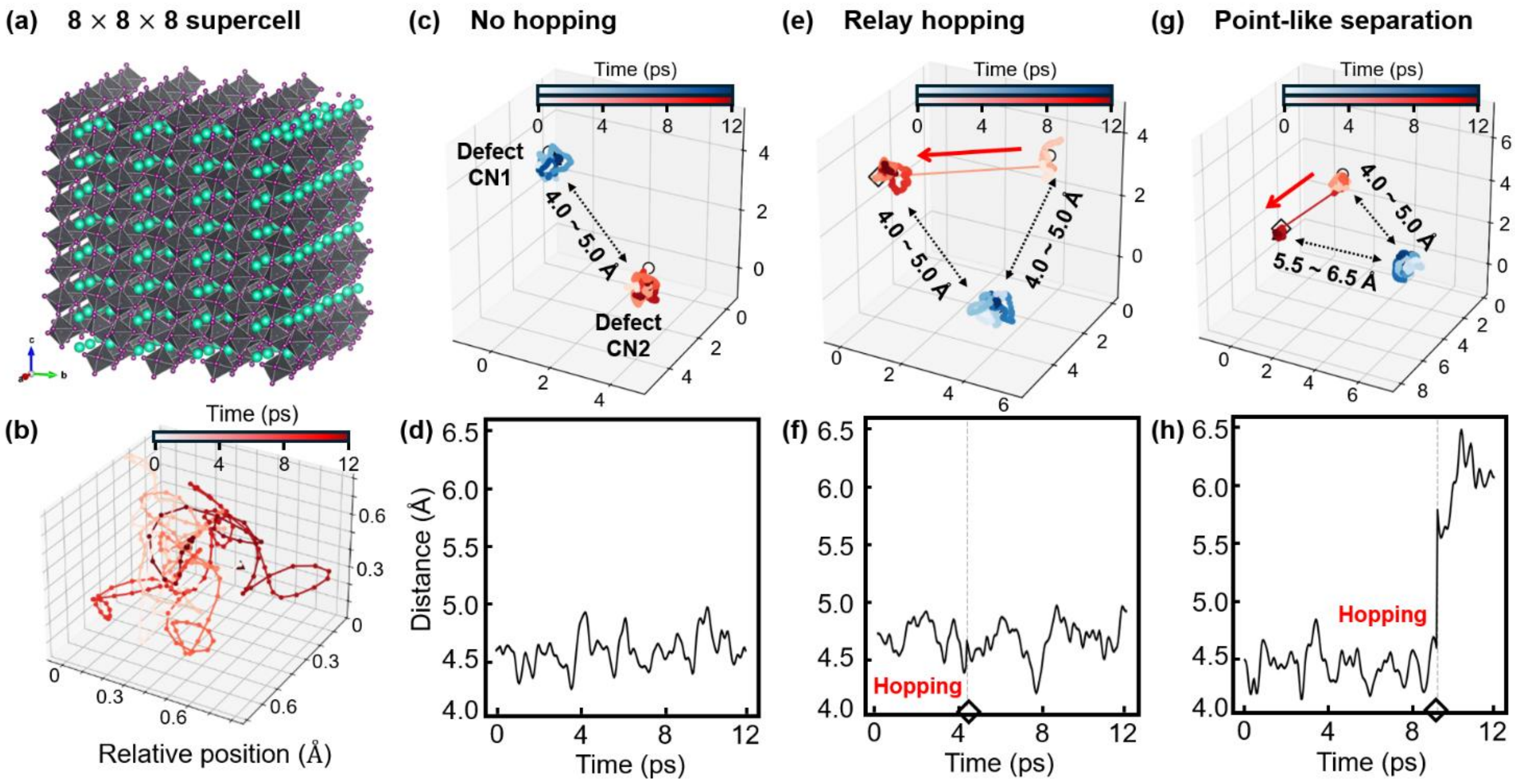

**Figure 2. Vacancy tracking and kinetic trajectory classification.** (a) 8 × 8 × 8 tilted supercell (comprising 2,560 atoms) employed for the MLMD simulations. (b) Temporal evolution of a representative single iodine-vacancy trajectory relative to its initial spatial position. (c, d) Three-dimensional spatial trajectory (c) and corresponding transient inter-defect distance (d) for a vacancy pair exhibiting no hopping. (e, f) Spatial trajectory (e) and time-dependent distance profile (f) illustrating relay hopping, which dynamically maintains an inter-defect distance of 4.0–5.0 Å. (g, h) Spatial trajectory (g) and corresponding distance evolution (h) demonstrating point-like separation, evidenced by an increased spatial separation to 5.5–6.5 Å. The diamonds on the horizontal axes in (f) and (h) explicitly denote the detection of discrete hopping events.

A discrete hopping event was registered when a tracked vacancy migrated from its initial bridge site to an adjacent one. By correlating these temporal hopping events with the inter-vacancy distance profiles, the trajectories were classified into three distinct modes: no hopping (**Figs. 2c** and **2d**), relay hopping (**Figs. 2e** and **2f**), and point-like separation (**Figs. 2g** and **2h**). Relay hopping was defined as a bridge-site reconfiguration wherein the two vacancies retain their nearest-neighbor proximity, maintaining a dynamic inter-defect distance of approximately 4.0–5.0 Å (**Figs. 2e** and **2f**). Conversely, point-like separation denotes a transition to spatially separated bridge sites, characterized by an abrupt increase in the inter-defect distance to 5.5–6.5 Å (**Figs. 2g** and **2h**). The fraction of trajectories for each hopping class was subsequently quantified by normalizing the assigned trajectory counts against the 96 evaluated configurations, with orientation-resolved fractions calculated using the respective subset populations.

## 4. Strain-dependent vacancy-defect pair reconfiguration

The statistical analysis of vacancy-pair reconfiguration events reveals a profound dependence on the applied isotropic and biaxial strains **(Fig. 3)**. Both deformation regimes exhibit a nonmonotonic, V-shaped trend in the total hopping fraction; reconfiguration kinetics are largely suppressed near equilibrium (0 % to −1 % strain) but become significantly activated under elevated magnitudes of both compression and tension. Crucially, the enhanced hopping fractions at the two strain extremes are governed by divergent kinetic pathways. Under −2% isotropic compression, point-like separation decisively dominates over relay hopping (50.0% versus 13.5%) **(Fig. 3a** and **3b)**. Conversely, under +2% isotropic tension, this kinetic preference inverts entirely, with relay hopping emerging as the predominant mechanism (59.4% versus 15.6%). A qualitatively analogous trend manifests under biaxial strain **(Fig. 3c** and **3d)**. At −2% biaxial compression, point-like separation continues to exceed relay hopping, although the kinetic contrast between the two modes is distinctly less pronounced than under isotropic compression. Under biaxial tension, relay hopping again dictates the reconfiguration dynamics.

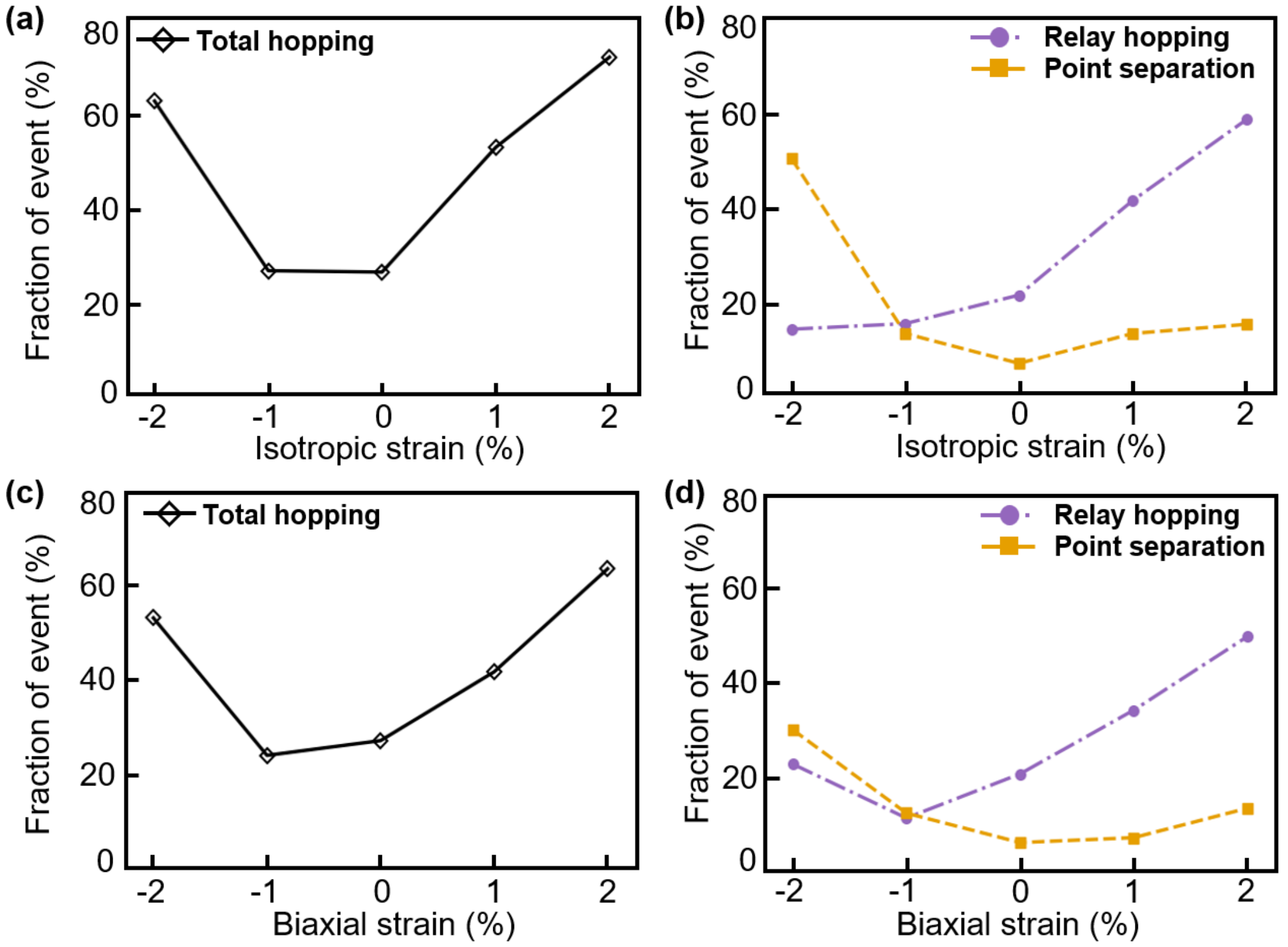

**Figure 3. Strain-dependent vacancy-pair reconfiguration dynamics.** Statistical evaluation of total hopping events (black solid lines) and their respective kinetic contributions from relay hopping (purple dash-dotted lines) and point-like separation (yellow dashed lines) under (a, b) isotropic and (c, d) biaxial strains. The occurrence fractions are evaluated across a strain range of −2% to +2%, with each data point strictly normalized against the 96 trajectories initialized from distinct representative vacancy-pair configurations simulated at the corresponding strain condition.

To characterize the structural environment sampled during these trajectories, we evaluated the temporal standard deviations of Pb-Pb and Cs-Cs distances within continuous 3ps time window containing no vacancy hopping (**Fig. 4**). Vacancy sites remain unchanged within each window, and both pair and separated configurations are included. The window-level results were averaged within each trajectory before calculating the means across trajectories. Under both strain modes, vacancy bridges exhibit greater Pb–Pb distance fluctuations than adjacent and bulk bridges, where the vacancy bridge denotes the Pb-Pb pair spanning an iodine vacancy site $V_I$ (**Fig. 4a** and **4c**). By comparison, the difference between defect-near and bulk Cs-Cs fluctuations is smaller, although both increase toward tensile strain (**Fig. 4b** and **4d**). Under isotropic strain, adjacent and bulk Pb–Pb fluctuations increase toward tension, whereas vacancy-bridge fluctuations decrease toward the unstrained condition and then increase modestly under tension (**Fig. 4a**). Consequently, the larger fluctuation contrast between vacancy and bulk bridges under strong isotropic compression indicates more pronounced local structural fluctuations relative to the bulk than under tension. This strain-dependent contrast is consistent with the greater prevalence of vacancy-pair separation under compression, suggesting a possible connection between local structural fluctuations and defect reconfiguration.

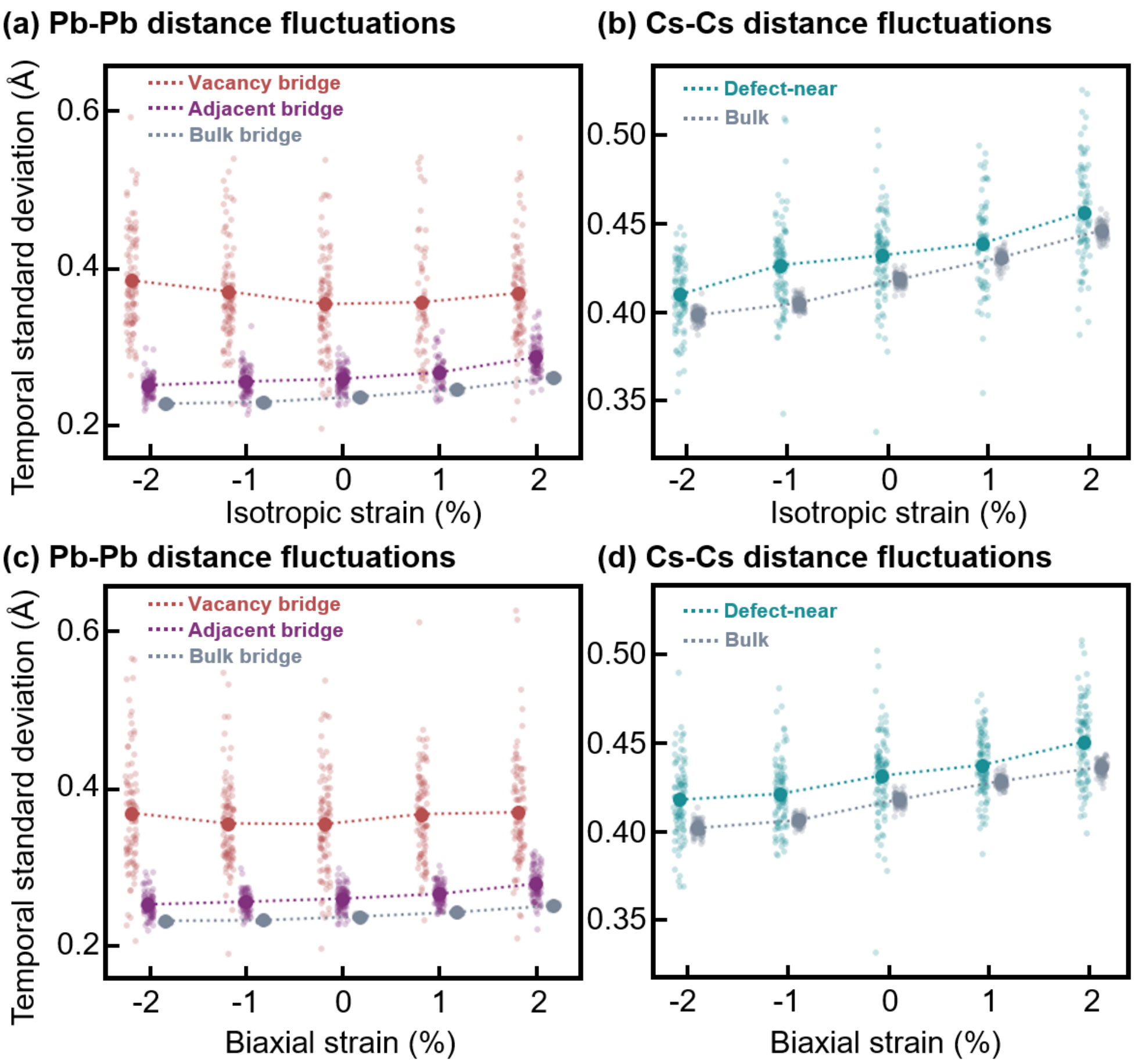

**Figure 4. Strain dependence of local and bulk framework distance fluctuations.** Temporal standard deviations of Pb-Pb and Cs-Cs distances under (a, b) isotropic and (c, d) biaxial strain. Thermal distance fluctuations were evaluated within continuous 3 ps time range containing no vacancy hopping, during which the vacancy sites remain unchanged. Pb-Pb fluctuations are compared among vacancy bridges, adjacent bridges, and bulk bridges; Cs-Cs fluctuations are compared between defect-near sites and inner bulk regions. Faint points represent individual trajectory averages over eligible windows, and filled markers connected by dotted lines indicate ensemble means.

Deconvoluting these kinetic responses based on the initial geometric orientation of the vacancy pairs reveals distinct strain-mode dependencies **(Fig. 5)**. Under isotropic strain, the A-type (out-of-plane), B-type (in-plane), and ambiguous-orientation pairs exhibit broadly analogous kinetic trajectories **(Figs. 5a** and **5b)**, indicating a weak orientation dependence. Specifically, point-like separation consistently dictates the dynamics under strong isotropic compression, whereas relay hopping governs under tension, regardless of the initial pair geometry. In stark contrast, biaxial strain induces a pronounced orientation-dependent kinetic bifurcation, most notably under −2% in-plane compression **(Figs. 5c** and **5d)**. Under this strong compression, the separation fraction is 48% for B-type pairs compared with 11% for A-type pairs, while the relay-hopping fraction 39% for A-type pairs compared with 14% for B-type pairs. Ambiguous-orientation pairs exhibit intermediate fractions for both event categories. At +2% biaxial tension, the A- and B-type fractions become nearly identical, at 53% and 52% for relay hopping and 16% and 10% for separation, respectively, with relay hopping more frequent than separation in all three orientation classes.

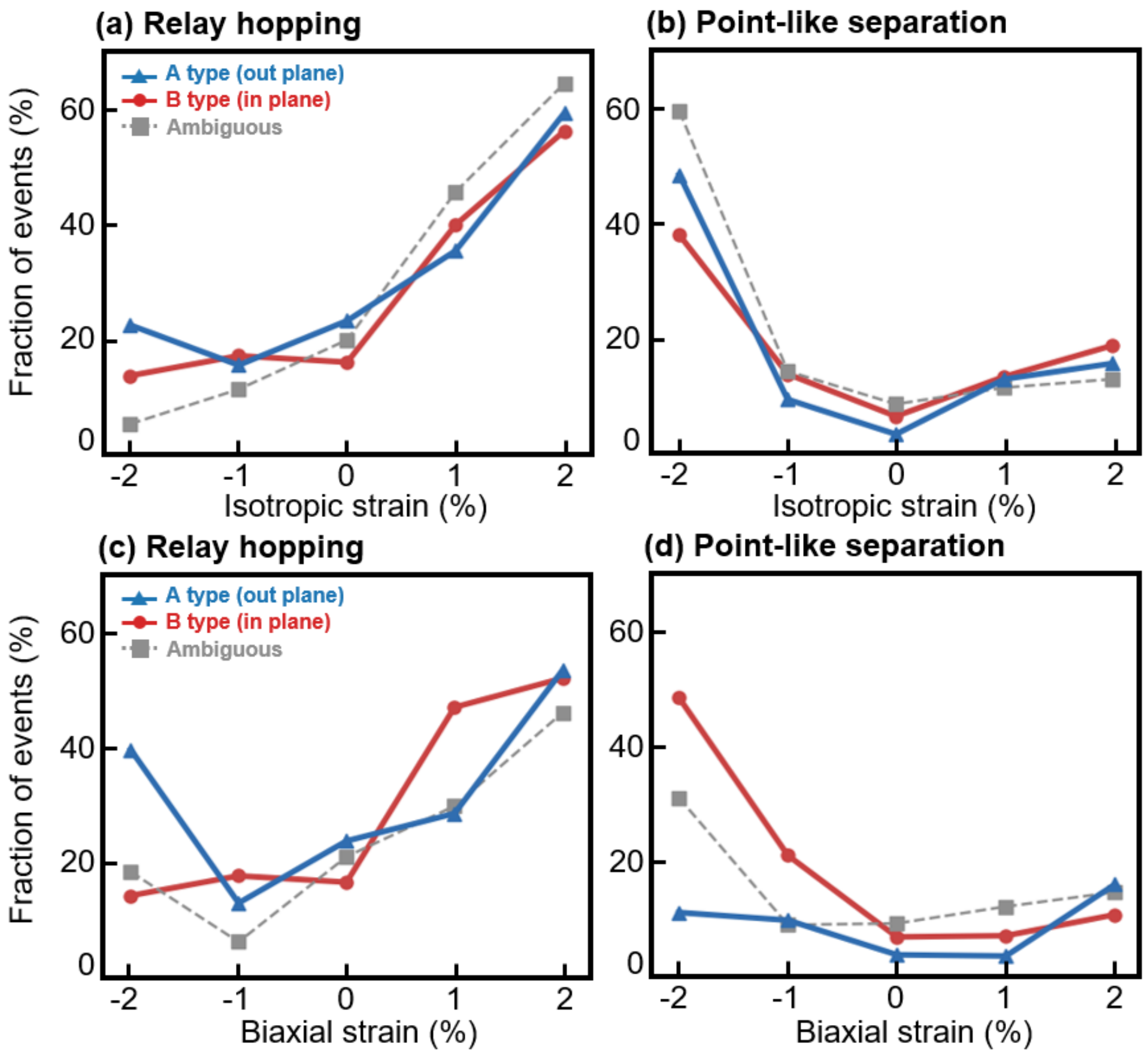

**Figure 5. Orientation-resolved vacancy-pair reconfiguration.** Relay-hopping and point-like-separation fractions for A-type, B-type, and ambiguous-orientation pairs under (a,b) isotropic and (c,d) biaxial strain. Each fraction is normalized by the valid trajectories within the corresponding orientation class. The orientation-specific denominators used to calculate the fractions are listed in Table S1 of the Supplemental Material.

Collectively, these results demonstrate that lattice strain modulates not only the overall occurrence of vacancy reconfiguration but also the relative prevalence of separation and relay hopping, with a pronounced dependence on initial pair orientation under biaxial compression. This orientation sensitivity is qualitatively consistent with previous reports of direction-dependent migration barriers in halide perovskites [10]. However, the present MLMD trajectories characterize short-time reconfiguration events rather than explicit activation barriers or macroscopic diffusivity. To complement these dynamical findings, we next evaluate the DFT interaction energies of representative vacancy pairs in the tilted lattice.

## 5. Vacancy-pair interaction energies

Following previous DFT studies of iodine-vacancy configurations in periodic $CsPbI_3$ supercells [11], we evaluate vacancy-pair interaction energies in a representative tilted lattice. To estimate the characteristic energy scale of vacancy-vacancy interactions, we calculated the interaction energies of representative in-plane and out-of-plane pairs within the tilted perovskite lattice. For two iodine vacancies initially located at sites $a$ and $b$, the interaction energy was defined as

$$E_{\text{int}}(\varepsilon) = E_{2V_I}^{a,b}(\varepsilon) + E_0(\varepsilon) - E_{V_I}^{a}(\varepsilon) - E_{V_I}^{b}(\varepsilon),$$

where $E_0$ is the energy of the pristine structure, $E_{V_I}^{a}(\varepsilon)$ and $E_{V_I}^{b}(\varepsilon)$ are the energies of structures containing a single iodine vacancy at sites $a$ and $b$ respectively, and $E_{2V_I}^{a,b}(\varepsilon)$ is the energy of the structure containing the vacancy pair. All four energies were calculated for charge-neutral defect conditions and evaluated at the same strain condition $\varepsilon$. With this thermodynamic convention, a negative $E_{\text{int}}$ indicates an attractive interaction, whereas a positive value indicates a repulsive interaction relative to the corresponding single-vacancy configurations.

The DFT calculations for the selected vacancy-pair configurations yielded positive interaction energies across all examined strain conditions, indicating weak repulsive interactions without evidence of strong static pair binding (**Fig. S2** in the Supplemental Material). To benchmark the SevenNet-0 interaction energies against DFT, we selected one unstrained in-plane vacancy pair and evaluated it in $2 \times 2 \times 2$ and $4 \times 4 \times 4$ supercells (**Table I**). For each supercell size, SevenNet-0 single-point energies were evaluated on the same DFT-relaxed pristine, single-vacancy, and vacancy-pair structures. Both methods yielded positive interaction energies for this pair at both supercell sizes. The absolute difference was 0.068 eV in the 2 × 2 × 2 supercell and 0.009 eV in the 4 × 4 × 4 supercell. This close quantitative agreement demonstrates that the MLFF can reproduce DFT interaction energetics for the examined configuration, providing support for extending the analysis to a broader range of vacancy-pair configurations. We then used MLFF with SevenNet-0 to survey 96 representative vacancy-pair configurations per strain condition in the same 4 × 4 × 4 supercell under isotropic (**Fig. 6a** and **6b**) and biaxial strain (**Fig. 6c** and **6d**). All examined configurations exhibited positive interaction energies, indicating consistently repulsive static interactions between neighboring iodine vacancies. The mean interaction energies generally decreased from compressive to tensile strain under both strain modes, with relatively small differences among pair orientations. Notably, even the minimum interaction energies remained positive across all examined strain conditions, confirming the repulsive nature of the interactions throughout the sampled configuration space. The selected pairs examined with DFT also yielded positive interaction energies, but their energies did not show the same systematic decrease across the isotropic strain series (Supplementary **Table S2**).

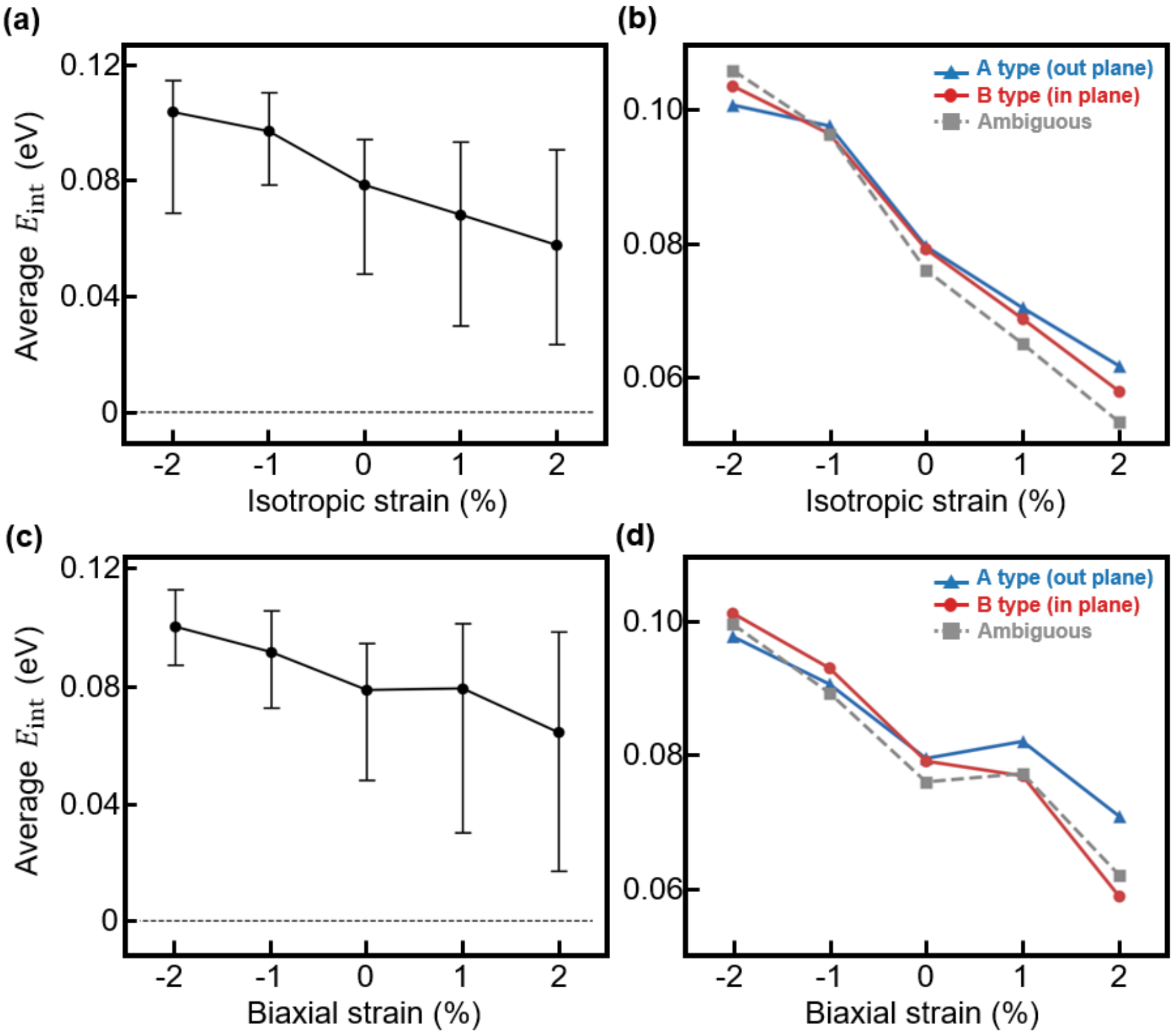

**Figure 6. Strain-dependent iodine-vacancy pair interaction energies calculated by MLFF using SevenNet-0.** (a) Mean interaction energies averaged over 96 vacancy-pair configurations under isotropic strain, with error bars indicating the minimum-to-maximum ranges. (b) Corresponding orientation-resolved mean interaction energies. (c) Mean interaction energies under biaxial strain. (d) Corresponding orientation-resolved mean interaction energies. Positive values indicate repulsive interactions.

These findings indicate that neighboring iodine vacancies are not energetically favored as static pairs in the examined configurations. Nevertheless, the distinct strain- and orientation-dependent kinetic preferences observed in MLMD cannot be explained by static interaction energies alone. The consistently repulsive interactions provide an energetic context for vacancy-pair reconfiguration, while dynamic lattice fluctuations may contribute to the observed competition between point-like separation and pair-preserving relay hopping.

**Table I.** DFT and MLFF interaction energies of a representative in-plane vacancy pair in unstrained $CsPbI_3$ for two supercell (2 × 2 × 2 and 4 × 4 × 4) sizes with lattice parameter $L$. SevenNet-0 single-point calculations used the corresponding DFT-relaxed structures. All results show repulsive interactions (positive $E_{\mathrm{int}}$).

| $L$ (Å) | DFT $E_{\mathrm{int}}$ (eV) | MLFF $E_{\mathrm{int}}$ (eV) | Absolute difference (eV) |
|---|---|---|---|
| 12.77 | +0.052 | +0.119 | 0.068 |
| 25.54 | +0.124 | +0.133 | 0.009 |

**6. Conclusions**

In this study, we investigated the dynamic reconfiguration of iodine-vacancy pairs in the tilted $CsPbI_3$ perovskite lattice under isotropic and biaxial strain using machine-learning molecular dynamics (MLMD) and density functional theory (DFT). The MLMD trajectories reveal a nonmonotonic dependence of the total hopping fraction on strain under both deformation modes. Strong isotropic compression predominantly favors point-like separation, whereas tensile strain promotes pair-preserving relay hopping. Notably, under −2% biaxial strain, separation was more frequent than relay hopping for initially in-plane pairs, whereas relay hopping was more frequent than separation for initially out-of-plane pairs.

Static DFT calculations yielded positive interaction energies for the selected iodine-vacancy pairs, and the SevenNet-0 survey likewise yielded positive values for all 96 representative pair configurations examined at each strain condition. Despite this static repulsion, neighboring vacancies could retain their nearest-neighbor relationship during relay hopping. Static interaction energies alone therefore cannot explain the observed dependence of pair separation and relay hopping on strain and initial pair orientation. This atomistic perspective may inform future strain-engineering strategies, including substrate selection and interfacial stress management, aimed at improving the operational stability of perovskite-based optoelectronic devices.

## Author Contributions

The manuscript was written through contributions of all authors. **Yonggoon Jeon** supervised the research project. **Donggeun Kang** conceived the idea and conducted the research; performed the DFT and MLMD calculations and data analyses, and wrote the manuscript. **Seongju Lee** assisted in organizing and presenting the research findings and provided valuable comments and discussion. **D. K.,** and Y.J. wrote the manuscript with the help of **S. L.**

**Competing interests**: The authors declare no competing interests.

## Acknowledgement

This work was supported by a National Research Foundation of Korea (NRF) grant funded by the Korea Government Ministry of Science and ICT (MSIT) (RS-2026-25483121). This work was also supported by 2025 research fund of Korea Military Academy (Hwarangdae Research Institute). (RN: 2025B1006)

## Data availability

The data used to support the findings of this study will be available from the corresponding author upon request after acceptance.

# Supplemental Material:

# Strain-dependent Dynamics of Interactions of Iodine-vacancy pairs in $CsPbI_3$: A DFT and Machine-Learning Molecular Dynamics Study

Donggeun Kang[1], Seongju Lee[1], Yonggoon Jeon[1,*]

[1]*Department of Physics and Chemistry, Korea Military Academy, Seoul, 01805, Republic of Korea*

*Corresponding author: 12-10011@kma.ac.kr

## S1.Benchmarking SevenNet-0 for iodine migration

We benchmarked the pretrained SevenNet-0 potential against density functional theory (DFT) along linear and curved iodine migration paths in tilted $CsPbI_3$. Both paths connected the same two DFT-relaxed iodine-vacancy configurations. For each path, we evaluated relative single-point energies and Cartesian atomic force components with both methods on the same eight interpolated configurations.

Along the linear path (**Fig. S1a**), SevenNet-0 reproduced the overall shape of the DFT energy profile, with an energy root-mean-square error (RMSE) of 0.134 eV and a force RMSE of 0.093 eV/Å (**Figs. S1b** and **S1c**). Along the curved path (**Fig. S1d**), the energy and force RMSEs were 0.127 eV and 0.053 eV/Å, respectively (**Figs. S1e** and **S1f**). SevenNet-0 reproduced the position of the energy maximum along both paths and underestimated its height by approximately 0.2 eV along the curved path.

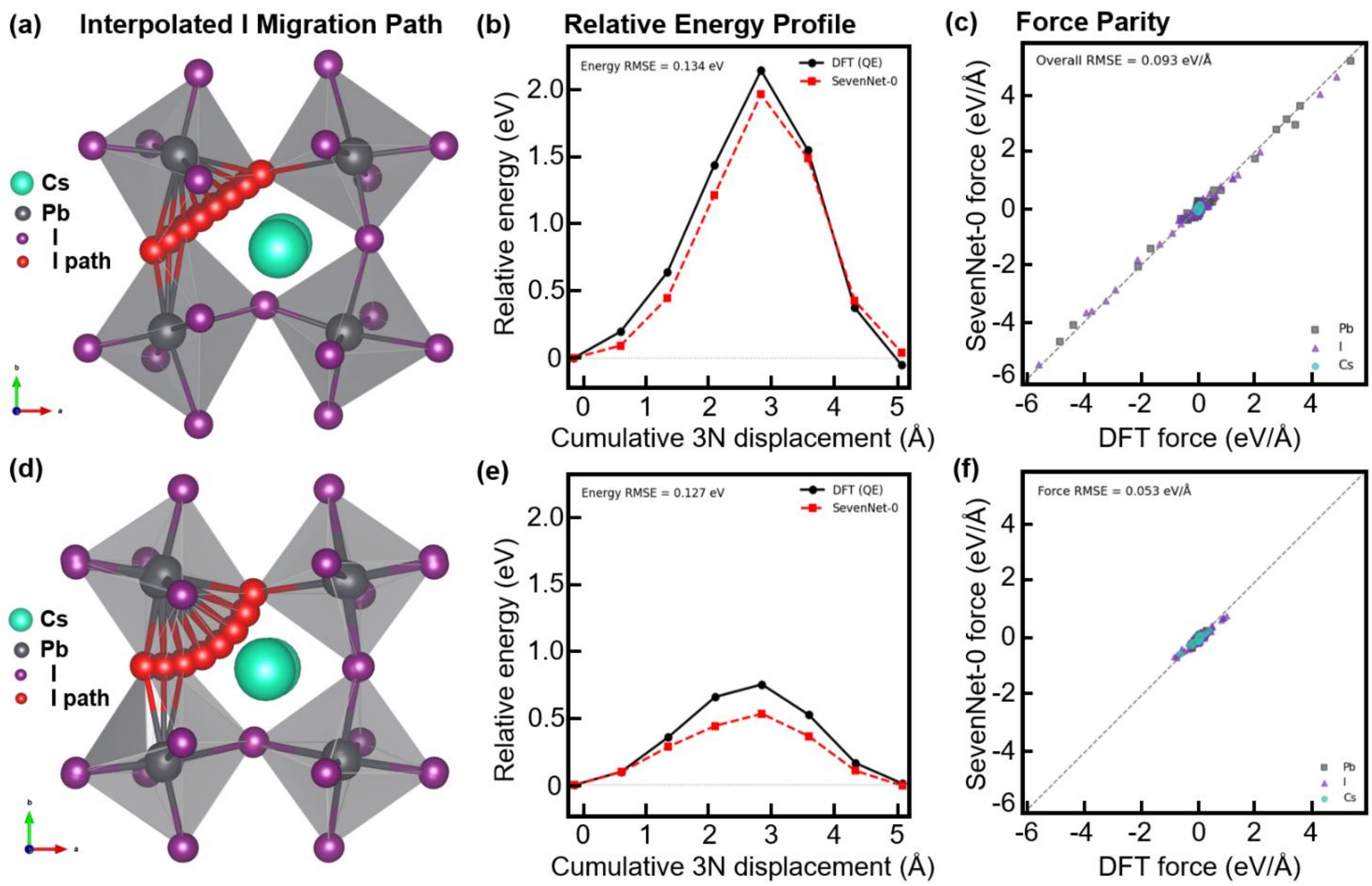

**Figure S1. Comparison of SevenNet-0 and DFT energies and forces along two iodine migration paths.** (a) Linear interpolation between two DFT-relaxed iodine-vacancy configurations. Red spheres mark successive positions of the migrating iodine atom. (b) Relative single-point energies calculated with DFT and SevenNet-0 for the same configurations along the linear path, plotted against the cumulative displacement in 3N-dimensional configuration space. The energy root-mean-square error (RMSE) is 0.134 eV. (c) Parity plot of Cartesian atomic force components along the linear path, with an overall RMSE of 0.093 eV/Å. (d) Curved interpolation between the same endpoint configurations, with the migrating iodine positions shown in red. (e) Relative single-point energies along the curved path, with an energy RMSE of 0.127 eV. (f) Force-component parity plot for the curved path, with an overall RMSE of 0.053 eV/Å. In (c) and (f), symbols distinguish Pb, I, and Cs, and the dashed lines indicate perfect agreement.

## S2. MLMD simulation and vacancy-tracking methodology

### A. Generation of initial vacancy-pair configurations

The initial iodine-vacancy pair configurations were generated from the strained pristine $CsPbI_3$ supercell by removing two iodine atoms located at neighboring sites within the $PbI_6$ octahedral network. Candidate vacancy pairs were required to share at least one neighboring Pb atom, with an initial I-I separation between 4.0 and 5.3 Å. To exclude opposite-site arrangements within the same octahedron, candidates with an I-Pb-I angle of 150° or greater were discarded. The Pb-I coordination network was constructed using a 4.0 Å neighbor cutoff, with sixfold Pb coordination required for candidate generation.

Symmetry operations of the strained pristine structure were identified using the spglib library, with a symmetry tolerance of $10^{-2}$ Å. Symmetry-equivalent vacancy-pair configurations were grouped into common orbits, and one representative structure was selected from each orbit. This procedure reduced redundant defect placements while retaining distinct local vacancy-pair configurations within the tilted lattice.

The initial orientation of each vacancy pair was characterized using the minimum-image displacement vector $\Delta\mathbf{r}$ connecting the two removed iodine sites. Its alignment with the out-of-plane direction was quantified as

$$\eta_z = |\cos\theta_z| = \frac{|\Delta z|}{|\Delta\mathbf{r}|}.$$

The initial configurations were classified as A-type (out-of-plane) for $\eta_z \geq 0.65$, B-type (in-plane) for $\eta_z \leq 0.35$, and ambiguous for $0.35 < \eta_z < 0.65$. This classification accounts for the nonideal pair orientations arising from octahedral tilting.

**B. Geometric definition of iodine-vacancy sites**

For the analysis of finite-temperature MD trajectories, each potential iodine-vacancy site was represented by a Pb-Pb bridge connecting two neighboring Pb atoms across an iodine site. Each bridge was assigned a unique identifier based on the corresponding Pb atom indices in the pristine lattice. The instantaneous position of bridge $i$ was defined as the midpoint between its two Pb atoms:

$$R_i(t) = \mathbf{r}_{\mathrm{Pb}_1}(t) + \frac{1}{2}\Delta\mathbf{r}_{\mathrm{Pb}_1\rightarrow\mathrm{Pb}_2}(t),$$

where $\Delta\mathbf{r}_{\mathrm{Pb}_1\rightarrow\mathrm{Pb}_2}$ is the Pb-Pb displacement vector evaluated under periodic boundary conditions using the minimum-image convention. Because the Pb positions fluctuate during MD, the bridge midpoint was recalculated at each saved frame. Thus, vacancy positions were determined with respect to the instantaneous local lattice geometry rather than fixed reference coordinates.

## C. Vacancy identification using bridge-occupancy scores

The occupancy of each candidate bridge was evaluated using the distance between its instantaneous midpoint and the nearest iodine atom, calculated under periodic boundary conditions using the minimum-image convention. This distance was defined as the **vacancy score** of the corresponding bridge. A larger score indicates that no iodine atom is located close to the bridge midpoint, making the bridge a stronger vacancy candidate. The candidate bridges were ranked in descending order of their scores, and two vacancy sites were selected for each frame. A minimum separation of 1.0 Å was imposed between selected candidate positions to avoid duplicate assignment of nearly identical sites.

To enable efficient tracking, the score evaluation was initially restricted to the bridge sites within the first neighboring shell of the previously identified vacancy sites. When the local detection score was insufficient, the search region was extended to the second neighboring shell. Importantly, the previous bridge IDs were used only to define the candidate search region; the vacancy sites themselves were reselected based on the instantaneous scores at each frame.

## D. Dynamic vacancy-pair identification and hopping detection

After identifying the two vacancy sites in each frame, their instantaneous separation was calculated as the minimum-image distance between the corresponding Pb-Pb bridge midpoints. The two vacancy sites were classified as a neighboring pair when their corresponding bridges shared a Pb atom and their separation did not exceed 5.5 Å. Otherwise, the configuration was classified as separated.

Vacancy migration was monitored by comparing the selected bridge IDs between consecutive saved frames. Changes in the selected bridge IDs were recorded as bridge-switch events, accompanied by the corresponding changes in vacancy positions and pair separation. To distinguish physically distinct vacancy rearrangements, the temporal bridge assignments and separation profiles were subsequently examined. The resulting trajectories were categorized into no hopping, pair-preserving relay hopping, and point-like separation, as illustrated in **Fig. 2** of the main text. Relay hopping denotes a vacancy-site rearrangement in which the two vacancies retain a neighboring-pair configuration, whereas point-like separation denotes a rearrangement that produces spatially separated vacancies. **Table S1** lists the numbers of trajectories in each initial orientation class used to normalize the relay-hopping and point-like-separation fractions.

**Table S1. Trajectory counts used as denominators for the orientation-resolved fractions in Fig. 5.** Counts are listed for A-type, B-type, and ambiguous initial orientations at each strain condition.

| Mode | Strain (%) | A-type $N_A$ | B-type $N_B$ | Ambiguous $N_M$ | Total orbit number |
|---|---|---|---|---|---|
| Isotropic | −2 | 31 | 28 | 37 | 96 |
| | −1 | 32 | 29 | 35 | 96 |
| | 0 | 30 | 31 | 35 | 96 |
| | +1 | 31 | 30 | 35 | 96 |
| | +2 | 33 | 32 | 31 | 96 |
| Biaxial | −2 | 28 | 29 | 39 | 96 |
| | −1 | 32 | 29 | 35 | 96 |
| | 0 | 30 | 31 | 35 | 96 |
| | +1 | 32 | 30 | 34 | 96 |
| | +2 | 32 | 29 | 35 | 96 |

## S3. Charge density difference (CDD) analysis

To further examine the electronic redistribution associated with neighboring iodine vacancies, we calculated the charge density difference (CDD) using the same defect configuration employed in the DFT interaction-energy analysis. The CDD was defined as

$$\Delta\rho(\mathbf{r}) = \rho_{2V_I}(\mathbf{r}) + \rho_{\text{pristine}}(\mathbf{r}) - \rho_{\text{V}_\text{I}^{(1)}}(\mathbf{r}) - \rho_{\text{V}_\text{I}^{(2)}}(\mathbf{r}),$$

where $\rho_{2V_I}$, $\rho_{\text{V}_\text{I}^{(1)}}$, $\rho_{\text{V}_\text{I}^{(2)}}$, and $\rho_{\text{pristine}}$ denote the electron densities of the vacancy pair, the two corresponding single-vacancy configurations, and the pristine lattice, respectively. All densities were evaluated using a consistent supercell and atomic reference geometry to isolate the electronic redistribution associated with vacancy-pair formation. As shown in **Fig. S2**, the CDD exhibits predominantly localized charge redistribution around the defect sites and neighboring atoms, without a clear signature of direct electronic interaction between the two vacancies, such as pronounced charge accumulation in the intervening region. This observation is qualitatively consistent with the weak vacancy-pair interaction energies obtained from DFT in **Table S2**. While the DFT calculations for selected configurations did not exhibit the systematic strain dependence observed in the MLFF results, both methods consistently yielded positive interaction energies, supporting the qualitative robustness of the repulsive interactions.

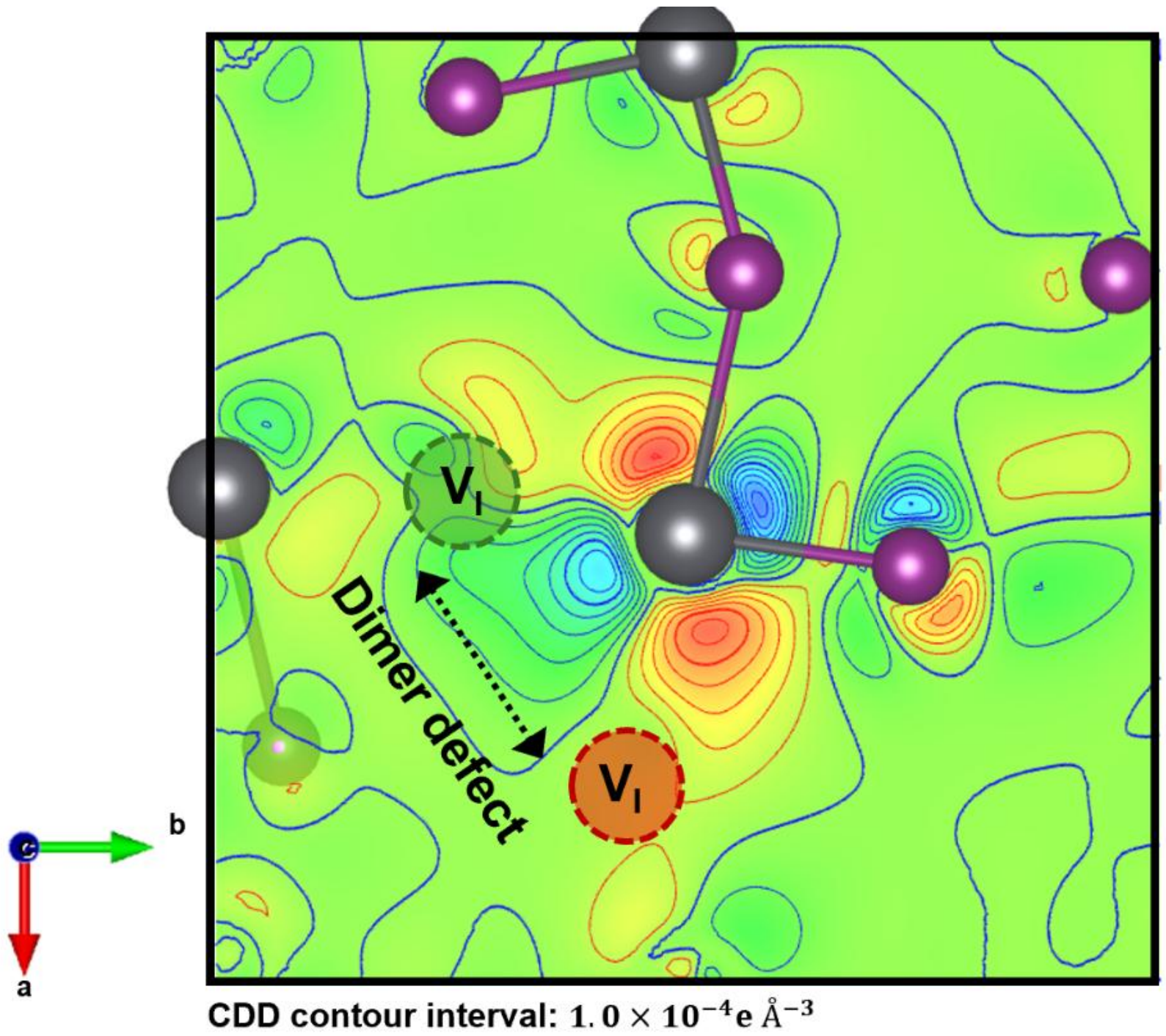

**Figure S2. Charge density difference (CDD) around a representative iodine-vacancy pair in tilted $CsPbI_3$ 2 × 2 × 2 supercell.** The CDD was calculated from the electron densities of the vacancy-pair, corresponding single-vacancy, and pristine configuration. Red and blue contours indicate charge accumulation (positive CDD) and depletion (negative CDD), respectively. Dashed circles indicate the two iodine-vacancy sites. The contour interval is $1.0 \times 10^{-4}$ e Å$^{-3}$.

**Table S2. DFT interaction energies ($E_{\mathrm{int}}$) of neighboring iodine vacancies in the tilted $CsPbI_3$ 2 × 2 × 2 supercell.** Comparative $E_{\mathrm{int}}$ analysis of out-of-plane A-type and in-plane B-type pairs under isotropic strain states (from −2 % to +2 %). Positive values indicate repulsive interactions.

| Isotropic strain (%) | A-type (eV) | B-type (eV) |
|---|---|---|
| −2.0 | +0.030 | +0.068 |
| −1.0 | +0.060 | +0.053 |
| 0 | +0.068 | +0.052 |
| +1.0 | +0.093 | +0.086 |
| +2.0 | +0.081 | +0.017 |